\documentclass[a4paper,amsmath]{revtex4}
\usepackage{amsmath, latexsym, amsfonts, amssymb, amsthm, amscd}
\usepackage{graphicx, color}

\usepackage[applemac]{inputenc}

\numberwithin{equation}{section}

\newtheorem{theorem}{Theorem}[section]
\newtheorem{lemma}[theorem]{Lemma}

\renewcommand{\tilde}{\widetilde}          
\DeclareMathSymbol{\leqslant}{\mathalpha}{AMSa}{"36} 
\DeclareMathSymbol{\geqslant}{\mathalpha}{AMSa}{"3E} 
\DeclareMathSymbol{\eset}{\mathalpha}{AMSb}{"3F}     
\renewcommand{\leq}{\;\leqslant\;}                   
\renewcommand{\geq}{\;\geqslant\;}                   

\newlength{\figurewidth}
\newlength{\figureheight}
\newcommand{\Z}{\mathbb{Z}}

\newcommand{\be}{\begin{equation}}
\newcommand{\ee}{\end{equation}}

\begin{document}

\title{Skew and implied leverage effect: smile dynamics revisited}

\author{Vincent Vargas}
\affiliation{ENS, 45 rue d'Ulm, 75005 Paris, France.}

\author{Tung-Lam Dao}
\affiliation{CFM, 23 rue de l'Universit\'e, 75007 Paris, France.}

\author{Jean-Philippe Bouchaud}
\affiliation{CFM, 23 rue de l'Universit\'e, 75007 Paris, France.\\ Ecole Polytechnique, 91120 Palaiseau, France.}


\begin{abstract} 
We revisit the ``Smile Dynamics'' problem, which consists in relating the implied leverage 
(i.e. the correlation of the at-the-money volatility with the returns of the underlying) and the skew of the option
smile. The ratio between these two quantities, called ``Skew-Stickiness Ratio'' (SSR) by Bergomi \cite{Ber2}, saturates to the value $2$ for 
linear models in the limit of small maturities, and converges to $1$ for long maturities. We show that for more general, non-linear models 
(such as the asymmetric GARCH model), Bergomi's result must be modified, and can be larger than $2$ for small maturities. The discrepancy comes from 
the fact that the volatility skew is, in general, different from the skewness of the underlying. 
We compare our theory with empirical results, using data both from option markets and from the underlying price series, for the S\&P500 and the DAX.
We find, among other things, that although both the implied leverage and the skew appear to be too strong on option markets, their
ratio is well explained by the theory. We observe that the SSR indeed becomes larger than $2$ for small maturities. 
\end{abstract}

\maketitle

\section{Introduction}

Among the best known ``stylized facts'' about option smiles are a) their skew, i.e. the fact that downstrike volatilities are generally higher than upstrike volatilities, reflecting the anticipated negative skewness of market moves and b) the ``implied leverage effect'', i.e. the tendency of at-the-money volatilities to increase when the underlying 
market goes down. A huge amount of effort has been devoted to building a theoretical framework that accounts quantatively for these features (see e.g. \cite{BaFo,BP,SABR,Bergomi}). Simple ``rules of thumb'' are used by market makers 
in order to relate these two effects. One of them is the ``sticky strike'' rule, which assumes that the implied volatility of an option only depends on its strike $K$. Assuming the smile to be locally 
linear around the money, one defines:
\begin{equation}\label{smile-def}
\sigma_{{\rm BS},T}(K) \approx \sigma_{{\rm{ATM}},T} \left[ 1 +  \text{Skew}_T \, \mathcal{M} \right], \quad {\mbox{where}} \quad \mathcal{M} = \frac{\ln (K/S)}{\sigma_{{\rm{ATM}},T} \sqrt{T}} \ll 1,
\end{equation}
and $\sigma_{{\rm{ATM}},T}=\sigma_{{\rm BS},T}(K=S)$ is the ``at-the-money'' implied volatility, $T$ the maturity, $\mathcal{M}$ is the rescaled moneyness, assumed to be small, and $\text{Skew}_T$ 
is the relative slope of the smile, that we will
call throughout the ``skew''. Assuming ``sticky strike'' as defined above therefore immediately leads to the following relation between the change of $\sigma_{{\rm{ATM}},T}$ and the skew:
\begin{equation}
\delta \sigma_{{\rm BS},T}(K) = 0 =  \delta \sigma_{{\rm{ATM}},T}  +  \text{Skew}_T \, \frac{\delta S}{S \sqrt{T}} \longrightarrow   \delta \sigma_{{\rm{ATM}},T} = - \text{Skew}_T \frac{\delta S}{S \sqrt{T}}.
\end{equation}
More generally, L. Bergomi \cite{Ber2} proposed to introduce the (maturity dependent) Skew-Stickiness Ratio (SSR) $R_T$ defined as:
\begin{equation} \label{SSR-def}
\delta \sigma_{{\rm{ATM}},T} := - R_T \, \text{Skew}_T \, \frac{\delta S}{S \sqrt{T}},
\end{equation}
with $R_T \equiv 1$ if the  above ``sticky strike'' rule holds, and $R_T \equiv 0$ for the so-called ``sticky delta'' rule, where the volatility only depends on the moneyness 
(and is thus trivially constant at the money, for $\mathcal{M}=0$). The definition of $R_T$ in \eqref{SSR-def} should be understood in the sense of a standard regression 
of $\delta \sigma_{{\rm{ATM}},T}$ against ${\delta S}/{S}$. 

Can these rules be given some theoretical foundation, and what value of $R_T$ should one expect? Recently, Bergomi \cite{Ber2} and Ciliberti-Bouchaud-Potters 
\cite{CBP,CBP-E} independently and using a slightly different framework, proposed
a theory for $R_T$ and compared the results with empirical data. Bergomi assumes a general linear model for the forward volatility dynamics and expands to lowest order in vol-of-vol, whereas
Ciliberti et al.
use a cumulant expansion for the smile. The two results coincide and provide the following expression for the SSR \footnote{Assuming a flat forward variance curve and time-translation invariance
of the underlying dynamics: see below, Eq. \eqref{SSR-lin}, for a more general formula.}:
\begin{equation}\label{SSR-eq}
R_T = \frac{\sum_{\ell=1}^{T} g_L(\ell)}{\sum_{\ell=1}^T (1 - \frac{\ell}{T}) g_L(\ell)},
\end{equation}
where $g_L(\ell)$ is the so-called {\it leverage correlation} function of the underlying price process \cite{BMP,BP,CBP}:
\be
g_L(\ell) = \frac{E[r_i r_{i+\ell}^2]}{\sigma^3},
\ee
where $r_i$ is the return at time $i$ and $\sigma^2=E[r_i^2]$ the average square volatility of the process. It is interesting to give an explicit expression for $R_T$ in the case where
$g_T(\ell)$ is a simple exponential function $-A \exp(-\ell/\tau)$ with a relaxation time $\tau$. This shape is actually not a bad approximation for major stock indices, with $A \sim 0.2$ and 
$\tau \sim 30-50$ days \cite{CBP}. The SSR then takes the following form:
\be
R_T \approx \frac{T (1 - e^{-T/\tau})}{T- \tau (1 - e^{-T/\tau})}, \qquad (\tau \gg 1),
\ee
which displays the following limits: $R_T \approx 2$ ($1 \ll T \ll \tau$) and $R_T \approx 1 + \tau/T$ ($T \gg \tau$). As shown in Bergomi \cite{Ber2}, these limiting values are in fact independent of 
$g_L(\ell)$, provided it decays fast enough for large $\ell$. Note however that for $\tau=50$ days and $T=250$ days (1 year of trading), $R_T \approx 1.25$ still substantially larger than unity. 

These results are interesting, but the framework within which they were obtained turns out to be restrictive, for several reasons. First, as shown in  \cite{DVCB}, the cumulant expansion for 
the smile (and hence the theoretical approximation for the skew $\text{Skew}_T$) is very inaccurate in practice. An alternative, general smile formula, without any assumption on the underlying model (except the existence of all moments of order $\leq 2$ for the returns) was derived in \cite{DVCB}. One of the salient features of this new formula is that the coefficients of the quadratic expansion involve low moments of the return distribution which do not necessarily coincide with the coefficients given by a standard cumulant expansion. Second, the class of linear models considered by Bergomi \cite{Ber2} and 
Bergomi \& Guyon \cite{BG} cannot handle the strong, non-linear leverage effect that seems to characterize stock index returns. The main purpose of this paper is to show that one can derive analytically the skew term of the expansion within a large class of non linear Gaussian models. We specialize our general results to the case of the totally assymetric GARCH model, which is believed to 
provide a good description of the (non-linear) leverage effect of stock indices. We finally compare our theoretical results with  empirical data and comment on the remaining discrepancies.

\section{Main theoretical results}

\subsection{The framework}

Similarly to \cite{Bergomi}, we choose to model the full forward variance curve $\lbrace v_i^{i+\ell}\rbrace_{\ell \geq 0}$ directly. This approach is general and flexible as it does not assume any shape for the 
instantaneous forward variance curve; in particular, one can either choose to calibrate the model on the market forward variance curve (using the options market) 
or assume some form for the forward variance curve using either historical data or an underlying variance model. We introduce a sequence $(\epsilon_i)_{i \in \Z}$ of
i.i.d. standard Gaussian variables and an arbitrary  function $f$ that satisfies $E[f(\epsilon_i) ]=0$. Let $\mathcal{F}_i= \sigma \lbrace  \epsilon_j; \: j \leq i  \rbrace$ be the relevant filtration.
We adopt the following framework for the (log) price $S_{i}$:
\begin{equation}\label{definition}
r_i:= \ln \frac{S_{i+1}}{S_i} =\sigma_i \: \epsilon_i, \quad v_{i+1}^{i+\ell}-v_i^{i+\ell} = \nu \lambda_i^{i+\ell}(\lbrace v_i^u \rbrace_{u \geq i}) f(\epsilon_i) 
\end{equation}
where $v_i^{i+\ell}= E\left[\sigma_{i+\ell}^2  | \mathcal{F}_{i-1} \right ]$ is the forward variance (with $\sigma_i^2 \equiv v_i^i$), $\nu$ is an expansion parameter (the vol of vol)
and the set of $\lambda_i^{i+\ell}(\lbrace v_i^u \rbrace_{u \geq i})$ are arbitrary functions that describe the coupling of the variance curve with the current residual return $\epsilon_i$, that 
may themselves depend on the current forward variance curve $\lbrace v_i^u \rbrace_{u \geq i}$.
The initial ($t=1$)
variance curve is $v_1^j$, $j \geq 1$, and the total expected variance up to maturity is $V_T = \sum_{j=1}^T v_1^j$. In the linear case, i.e. $f(x)=x$, 
we recover (in a discrete setting) the framework of Bergomi \& Guyon \cite{BG}. In the sequel, we will restrict to a first order expansion in $\nu$ as we focus in this paper on the skew term in the smile expansion \eqref{smile-def}. Within this approximation, the functions $\lambda_i^{i+\ell}$ are functions of the initial (deterministic) curve $v_1^j$. We will make this dependence implicit, as this lightens the notations.

Note that since the $\epsilon_i$ are i.i.d Gaussian variables with $E[f(\epsilon_i) ]=0$, one deduces the following equality, that relates $\lambda_i^{i+\ell}$ to the leverage correlation function:

\be \label{eq-lambda}
E[r_i r_{i+\ell}^2] = \nu \sqrt{v_i^i} \lambda_i^{i+\ell} E[f'(\epsilon_i)].
\ee


\subsection{The smile formula}

We get the following smile formula at order 1 in $\nu$ (see proof in the appendix):
\begin{equation}
\sigma_{{\rm BS},T}= \sqrt{\frac{V_T}{T}}+  \frac{\nu}{2\sqrt{V_T T}}  \sum_{i=1}^T \sum_{j=1}^{i-1} \lambda_j^i  E \left[ f\left( \frac{\sqrt{v_1^j}}{V_T}
\left( \ln\left(\frac{K}{S}\right) +\frac{V_T}{2} \right) +Y_j\right)\right]
\end{equation}
where $Y_j$ a centered gaussian variable of variance $E[Y_j^2]=1- v_1^j/V_T$. 

We therefore get the following smile expansion at order $1$ in $\nu$ and in the modified rescaled moneyness $\mathcal{M}: = (\ln(K/S)+\frac{V_T}{2})/\sigma_{{\rm{ATM}},T} \sqrt{T}$:
\begin{equation}\label{smile-def2}
\sigma_{{\rm BS},T}= \sigma_{{\rm{ATM}},T} \left[1 + \text{Skew}_T \mathcal{M}\right], \qquad \sigma_{{\rm{ATM}},T} \equiv \sqrt{\frac{V_T}{T}}, 
\end{equation}
where the skew is given by the general expression 
\begin{equation}
\text{Skew}_T= \frac{\nu}{2 V_T^{3/2}}  \sum_{i=1}^T \sum_{j=1}^{i-1}  \sqrt{v_1^j} \lambda_j^i  E[f'(Y_j)] 
\end{equation}

Note that our method in fact enables one to derive a smile formula to quadratic order in $\mathcal{M}$ and to second order in $\nu$. For the sake of simplicity, we do not write the corresponding (cumbersome) expressions here. However, in the linear case $f(x)=x$, we recover exactly the formulae of Bergomi-Guyon \cite{BG} established in a continuous setting, by working on a time step $\delta t$ and then taking the limit as $\delta t \to 0$.        

\subsection{Skew and Skewness}

Recall that using a standard cumulant expansion one derives the following smile formula (\cite{BaFo,BCP,BP}): 
\begin{equation}
\sigma_{{\rm BS}}= \sigma_{{\rm{ATM}},T} \left[1 + \frac{\mathcal{S}_T}{6} \mathcal{M}\right], \qquad \sigma_{{\rm{ATM}},T} \equiv \sqrt{\frac{V_T}{T}}, 
\end{equation}
where $\mathcal{S}_T$ is the skewness of  $\ln (S_{T}/S_1)$, i.e. the return between now and maturity. Within the present framework, the skewness at order 1 in $\nu$
can be computed as:
\begin{equation}\label{skewness}
\frac{\mathcal{S}_T}{6}= \frac{\nu}{2 V_T^{3/2}} \left( \sum_{i=2}^T \sum_{j=1}^{i-1}  \sqrt{v_1^j} \lambda_j^i \right) E[f'(\epsilon)], 
\end{equation}
which is close to, but different from, the formula above for the skew $\text{Skew}_T$. However, for a linear model where $f'(x) \equiv 1$, the two formulas exactly coincide.
To first order in $\nu$, the skewness of the returns and the skew of the smile are therefore identical for a general linear model:
\begin{equation}
f(x) = x \longrightarrow \text{Skew}_T \equiv \frac{\mathcal{S}_T}{6}.
\end{equation}


\subsection{The implied leverage coefficient and the SSR}

One can also readily compute the implied leverage within our general non-linear model. The difference between tomorrow's volatility and today's volatility 
for {\rm{ATM}} options of maturity $T$ is given by:
\be
\frac1T \delta V_T = \frac1T \left[\sum_{j=2}^{T+1} v_2^j - \sum_{j=1}^{T} v_1^j \right] = \frac1T (v_1^{T+1} - v_1^1) + \frac{\nu f(\epsilon_1)}{T} \left[\sum_{j=2}^{T+1} \lambda_1^j \right]  
\ee
Using $\delta V_T = 2T \sigma_{{\rm{ATM}},T}\, \delta \sigma_{{\rm{ATM}},T}$, $E[r_1 f(\epsilon_1)] \equiv \sigma_1 E[f'(\epsilon_1)]$ and integration by parts for the Gaussian variable $\epsilon_1$ \footnote{We recall that $\sigma_1 \equiv \sqrt{v_1^1}$.}, 
we finally deduce a formula for the implied leverage $\gamma_T$, i.e. the correlation between the return and the change of implied {\rm{ATM}} volatility as:
\be
\gamma_T:= \frac{E[\delta \sigma_{{\rm{ATM}},T} \cdot r_1]}{E[r_1^2]} = \frac{\nu E[f'(\epsilon_1)]}{2\sqrt{T \sigma_1^2 V_T}} \left[\sum_{j=2}^{T+1} \lambda_1^j \right].
\ee
where we have used $E(\epsilon_1)=0$. Here, we have neglected any drift effect, which is reasonable for option pricing in a risk neutral framework. 

The average SSR is defined, consistently with Eq. (\ref{SSR-def}), as:
\be
{R_T} := \frac{\gamma_T \sqrt{T}}{\text{Skew}_T}.
\ee

In a {\it linear model} where skew and skewness are identical (up to a factor $6$), the final expression for the SSR is therefore given by:
\be \label{SSR-lin}
\left.\widehat{R}_T\right|_{lin.} = \frac{V_T}{\sigma_1} \frac{\sum_{j=2}^{T+1} \lambda_1^j}{\sum_{i=2}^T \sum_{j=1}^{i-1}  \sqrt{v_1^j} \lambda_j^i},
\ee
which is precisely Bergomi's result in a discrete time setting \cite{Ber2}. Note that for a {\it flat forward variance} curve and for a {\it time-translation invariant model}, one has
$v_1^j = v_1^1 \equiv \sigma_1^2$, $V_T = T v_1^1$, and $\lambda_j^i \propto g_L(j-i)$ where $g_L$ is the leverage correlation function introduced above.
In this case, one recovers exactly Eq. (\ref{SSR-eq}) above.

However, in the general case, ${R}_T$ is not given by the above expression, but is corrected by a factor that accounts for the difference between the skew and the
skewness:
\be \label{SSR-final}
{R_T} = \left.\widehat{R}_T\right|_{lin.} \times \frac{\mathcal{S}_T/6}{\text{Skew}_T}.
\ee
We will study below this correction factor, both within the asymmetric GARCH model and using empirical data.  

Note finally that using Eq. (\ref{eq-lambda}), the implied leverage  can alternatively be re-written in terms of the leverage correlation function as:
\be \label{imp-leveragefirst}
\gamma_T = \frac{1}{2\sqrt{T \sigma_1^4 V_T}} \left[\sum_{j=2}^{T+1} E[r_1 r_{j}^2] \right],
\ee
or, for a flat forward variance curve; 
\be \label{imp-leverage}
\gamma_T = \frac{1}{2T \sigma_1^3} \left[\sum_{j=2}^{T+1} E[r_1 r_{j}^2] \right].
\ee

In fact, expressions \eqref{imp-leveragefirst}  and \eqref{imp-leverage} for $\gamma_T$ are quite general and valid in a much more general setting than the one we consider here; indeed, they do not rely on any modelling assumption for the returns $r_i$.

\section{The asymmetric GARCH model}

In order to give some flesh to the above formulae in the context of an empirically relevant, non-linear model for price changes, we consider
the following so-called fully asymmetric GARCH model:
\begin{equation}
r_i=\sigma_i \epsilon_i, \quad \sigma_{i+1}^2=v_0^2+\rho(\sigma_{i}^2-v_0^2)+\nu \sigma_i^2\left(\epsilon_i^2 1_{\epsilon_i <0 }-\frac{1}{2}\right) 
\end{equation}
where  $(\epsilon_i)_{i \in \Z}$ are i.i.d. standard Gaussian random variables. In the notation above, one has $f(x)=x^2 1_{x < 0} - 1/2$.

We set $\sigma_{i}^2=v_0^2 (1+\mathcal{X}_i)$, so that the above is equivalent to the following recursion 
\begin{equation}
\mathcal{X}_{i+1}= \rho \mathcal{X}_i +\nu(1+ \mathcal{X}_i) ( \epsilon_i^2 1_{\epsilon_i <0 }-\frac{1}{2}  ) 
\end{equation}

By iterating the above expression, we get the following exact expression for $\sigma_i^2$ ($i \geq 2$):
\begin{equation}
\sigma_i^2= v_0^2 (1+\rho^{i-1}\mathcal{X}_1)+\nu \sum_{j=1}^{i-1} \rho^{i-1-j} \sigma_j^2 \left(    \epsilon_j^2 1_{\epsilon_j <0 }-\frac{1}{2}   \right)
\end{equation}

Therefore, to first order in $\nu$, we get, for arbitrary $j$ and $i$ with $j-i \geq 1$:
\begin{equation}
\sigma_j^2= v_0^2 (1+\rho^{j-i}\mathcal{X}_i)+
\nu V_0^2 \sum_{k=i}^{j-1} \rho^{j-1-k} (1+\rho^{k-i} \mathcal{X}_i ) \left(    \epsilon_k^2 1_{\epsilon_k <0 }-\frac{1}{2}   \right)
\end{equation}
which leads to the following expression for the forward variance curve \footnote{We note that $E\left[  \epsilon^2 1_{\epsilon <0 }-\frac{1}{2} \right] = 0$}
\begin{equation}
v_i^j = E[  \sigma_{j}^2  | \mathcal{X}_i  ]= v_0^2 (1+\rho^{j-i} \mathcal{X}_i )
\end{equation}

To first order in $\nu$, we also obtain:
\begin{align*} 
v_{i+1}^j -v_i^j  &= \nu v_0^2 \rho^{j-i-1} (1+\mathcal{X}_i ) \left(\epsilon_i^2 1_{\epsilon_i <0 }-\frac{1}{2} \right)  \\
&= \frac{\nu}{\rho} (v_0^2(\rho^{j-i}-1)+ v_i^j  ) \left(    \epsilon_i^2 1_{\epsilon_i <0 }-\frac{1}{2} \right) \\
\end{align*}
Therefore, we finally obtain for the $\lambda_i^j$ (which indeed explicitly depend on the forward rate):
\begin{equation}
\lambda_i^j = \frac{1}{\rho} (v_0^2(\rho^{j-i}-1)+ v_i^j)
\end{equation}

Finally, since $f'(x)= 2x 1_{x<0}$, we get the following expression for the skew and the skewness 
\footnote{We remind that for Gaussian variables $E[2Y 1_{Y<0}] = \sqrt{2/\pi}\sqrt{E[Y^2]}$. Note in passing that for 
a symmetric GARCH model, $E[f']=0$ and the skew/skewness disappear.} :
\begin{align*}
\text{Skew}_T & = - \sqrt{\frac{2}{\pi}}  \frac{\nu}{2 V_T^{3/2}}  \sum_{j=2}^T \sum_{i=1}^{j-1}  \sqrt{v_1^i} \lambda_i^j  \sqrt{1-\frac{v_1^i}{V_T}}  \\
&   = - \sqrt{\frac{2}{\pi}}  \frac{\nu v_0^3}{2 V_T^{3/2}}  \sum_{j=2}^T \sum_{i=1}^{j-1}  \sqrt{1+\rho^{i-1} \mathcal{X}_1} ( \rho^{j-i-1}+ \rho^{j-2}   \mathcal{X}_1 ) 
\sqrt{1-\frac{v_1^i}{V_T}}
\end{align*}
and
\begin{align*}
\frac{\mathcal{S}_T}{6}  & = - \sqrt{\frac{2}{\pi}}  \frac{\nu}{2 V_T^{3/2}}  \sum_{j=2}^T \sum_{i=1}^{j-1}  \sqrt{v_1^i} \lambda_i^j   \\
&  = - \sqrt{\frac{2}{\pi}}  \frac{\nu v_0^3}{2 V_T^{3/2}}  \sum_{j=2}^T \sum_{i=1}^{j-1}  \sqrt{1+\rho^{i-1} \mathcal{X}_1  } ( \rho^{j-i-1}+ \rho^{j-2}   \mathcal{X}_1 )   
\end{align*}
Here, the variance $V_T$ is given by:
\begin{align*}
V_T &= \sum_{i=1}^T v_1^i = v_0^2\left(T+  \frac{1-\rho^T}{1-\rho} \mathcal{X}_1 \right),
\end{align*}
therefore leading to the following skewness/skew ratio:
\begin{equation}\label{ratio}
\frac{\mathcal{S}_T/6}{ \text{Skew}_T}= \frac{ \sum_{j=2}^T \sum_{i=1}^{j-1}  \sqrt{1+\rho^{i-1} \mathcal{X}_1  } ( \rho^{j-i-1}+ \rho^{j-2}   \mathcal{X}_1 )}{ \sum_{j=2}^T \sum_{i=1}^{j-1}  \sqrt{1+\rho^{i-1} \mathcal{X}_1  } ( \rho^{j-i-1}+ \rho^{j-2}   \mathcal{X}_1 )    \sqrt{1-\frac{1+\rho^{i-1}   \mathcal{X}_1  }{T+  \frac{1-\rho^T}{1-\rho} \mathcal{X}_1 }}}
\end{equation} 
This expression drastically simplifies when the initial volatility is equal to the average volatility, i.e. $\mathcal{X}_1 = 0$. In this case, one simply 
obtains:
\begin{equation}\label{ratioS}
\frac{\mathcal{S}_T/6}{\text{Skew}_T}=\sqrt{\frac{T}{T-1}}, 
\end{equation} 
which is equal to $2$ for $T=2$ and tends to unity when $T \to \infty$. We see clearly that in this model, the skew is systematically smaller than its
third cumulant estimate, i.e. the skewness. Dividing by the skewness instead of the skew therefore leads to an underestimate of the ``true'' SSR. 
Finally, the implied leverage is given by:
\begin{equation}
\gamma_T \approx - \frac{\nu \sqrt{1+\mathcal{X}_1 }}{\sqrt{2 \pi} \sqrt{T (T+  \frac{1-\rho^T}{1-\rho} \mathcal{X}_1 )}}\frac{1-\rho^T}{1-\rho}. 
\end{equation}


%

\section{Data analysis: skew, skewness and SSR}

The central result of our paper is given by Eqs. (\ref{SSR-lin}, \ref{SSR-final}), that relates Bergomi's Skew-Stickiness Ratio (SSR) $R_T$ 
to empirically measurable quantities. The three questions 
we want to address here are:
\begin{enumerate}
\item How well does our central result Eq. (\ref{SSR-final}) account for the SSR of index option markets?
\item How strong is the correction factor $\mathcal{S}_T/6\, \mathrm{Skew}_T$, induced by non-linear effects? 
\item How well are these features reproduced by the (non-linear) asymmetric GARCH model investigated in the above section?
\end{enumerate}

In order to discuss these issues, we need data both from the option markets and from the underlying contract. 
We have focused on two markets, S\&P 500 index and DAX, for which we have full information on both the underlying 
and on the option smiles for various maturities. Our data set runs from 2000 to 2013 for the S\&P 500 and from 2002 to 2013 for the DAX.

We extract from the data various statistical quantities. 
\begin{enumerate}
\item From the historical returns $r_i$ of the underlying index, we measure:
\begin{itemize} 
\item The leverage correlation function $g_L(\ell)$, obtained as a time average of the ratio $r_i r_{i+\ell}^2/\sigma_i^3$, where $\sigma_i^2$ is a 20 day 
exponential moving average (EMA) of the past squared 
returns. 
\item A low moment estimator of the skewness of the distribution of returns over $T$ days, defined as:
\begin{equation}\label{beta}
\beta_T = \sqrt{\frac{\pi}{2}}\left[1 - 2P(\tilde r_T>0)\right],
\end{equation}
where $\tilde r_T$ is the {\it detrended} T-day return and $P(\tilde r_T>0)$ is the probability that $\tilde r_T$ is positive. We determined the local drift using an EMA filter with timescale $T=1000$ days.
\end{itemize}

It turns out that the skew of the smile, $\text{Skew}_T$, should be on average equal to $\beta_T$ for fairly priced options \cite{DVCB}. 
Moreover, the standard skewness of the returns, $\mathcal{S}_T$, can be obtained from the 
leverage correlation function as \cite{BP}   
\begin{equation}\label{skewness}
\mathcal{S}_T = \frac{\zeta_1}{\sqrt{T}} + \frac{3}{\sqrt{T}}\sum_{\ell=1}^T\left(1-\frac{\ell}{T}\right) g_L(\ell)
\end{equation}
where $\zeta_1$ is the skewness of daily returns. 

\item From option prices we extract the volatility smile for different moneyness and maturities, which allows us to measure:
\begin{itemize} 
\item The skew of the smile $\text{Skew}_T$, defined from Eq. \eqref{smile-def}, that we average over the whole time period, for a set of 
fixed maturities.
\item The implied leverage coefficient $\gamma_T$, measured as the regression coefficient of the changes of {\rm{ATM}} implied vol on the returns of the underlying.
\item Finally, Bergomi's SSR is obtained as the time-averaged local SS Ratio, measured as:
\begin{equation}
\widehat{\mathcal{R}}_T(t;M) = M \frac{\sum_{i=t-M}^{t}\left(\sigma_{{\rm{ATM}},T}(i+1)-\sigma_{{\rm{ATM}},T}(i)\right) \times r_i }
{\sum_{i=t-M}^{t} \text{Skew}_T(i) \times  \sum_{i=t-M}^{t} r_i^2}
\end{equation}
where $M=50$ is the size of the moving average window. Our empirical estimate of the SSR is then obtained as:
\begin{equation}
{\mathcal{R}}_T = \langle\widehat{\mathcal{R}}_T(t;M)\rangle_t. 
\end{equation}
\end{itemize}
\end{enumerate}

In order to compare $\gamma_T$ and ${\mathcal{R}}_T$ with theoretical estimates, we furthermore assume that the underlying process is time-translation
invariant and that the forward variance is flat on average, which allows us to obtain $\gamma_T$ from  Eq. \eqref{imp-leverage} \cite{CBP-E}:
\be 
\label{imp-leverage-2}
\gamma_T^{th.} \approx \frac{1}{2T} \sum_{\ell=1}^T g_L(\ell).
\ee

Our results are summarized in three figures, each showing data for the S\&P 500 (left) and DAX (right).
\begin{itemize}
\item In Fig. 1-a,b, we show as a function of the maturity $T$ the unconditional skew $\text{Skew}_T$ and unconditional implied leverage $\gamma_T$, 
both extracted from option data, which we compare with their theoretical estimates, 
$\beta_T$ and $\gamma_T^{th.}$, obtained using the historical returns of the underlying. We conclude that (a) the 
options skews on the S\&P index are  stronger than predicted by $\beta_T$, but match very well for the DAX; 
(b) the implied leverage $\gamma_T$ of S\&P options is well estimated for short maturies, but systematically too strong for larger maturities, 
as observed in \cite{CBP-E}. For the DAX, on the other hand, the implied leverage $\gamma_T$ appears to be
too weak for short maturities and too strong for large maturities.
\item In Fig. 2-a,b, we show Bergomi's SSR $R_T$ as a function of maturity, together with our theoretical estimate $R_T^{th.}=\gamma_T^{th.}\sqrt{T}/\beta_T$, 
based either on the historical returns (using Eq. \ref{imp-leverage-2}) or on the predictions based on the asymmetric GARCH model, itself
calibrated on historical returns \footnote{The parameters of the GARCH model are found to be $\rho = 0.988, \nu = 0.123$ and $v_0 = 0.179$ for S\&P500 and
$\rho = 0.9856, \nu = 0.133$ and $v_0 = 0.207$ for DAX. The characteristic time for volatility relaxation within this GARCH description is thus around
$13$ days.}. We find that the 
overall agreement is quite good, in particular with the predictions of the asymmetric GARCH model which are less noisy than the direct estimate based
on historical returns. Therefore, although both the implied leverage $\gamma_T$ and the skew $\text{Skew}_T$ appear to be too strong on option markets, their
ratio is about right! Still, observe that $R_T$ clearly becomes {\it larger} than the asymptotic Bergomi value $2$ for small maturities. This is due 
to the correction factor $(\mathcal{S}_T/6)/\text{Skew}_T > 1$ that appears in Eq. (\ref{SSR-final}), to which we turn next.
\item In Fig 3-a,b, we now plot the expected value of the correction factor $(\mathcal{S}_T/6)/\text{Skew}_T$ as a function of maturity, again using either a direct estimate based 
on Eqs. \eqref{beta}, \eqref{skewness}, or on the prediction of the asymmetric GARCH model with empirically calibrated parameters. We see that in the case of the S\&P 500 
the value of this ratio is overall not very well captured by the GARCH model, which underestimates this ratio, as already noticed in \cite{DVCB}. One of the known weaknesses of the 
GARCH model is that it fails to capture the long-range memory of the volatility process. This might explain part of the discrepancy seen here. In the case of the DAX, the ratio
is much closer to unity (except for small maturities where it is below $1$), suggesting that non-linear effects are weaker in this case. 
\end{itemize}

\begin{figure}[tbph]
\begin{center}
	\includegraphics[width = 0.6\figurewidth,height = 0.6\figureheight]{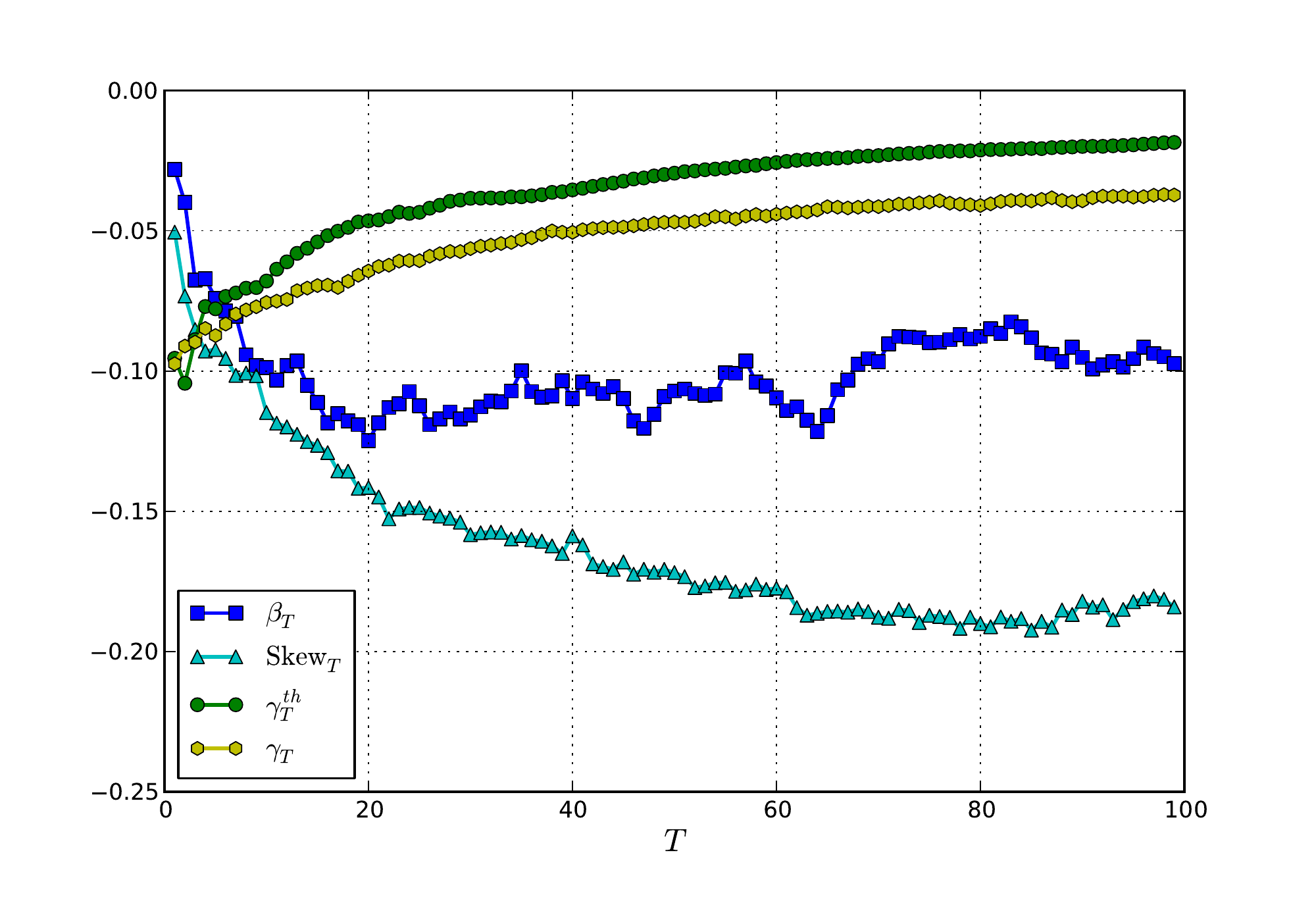}
	\includegraphics[width = 0.6\figurewidth,height = 0.6\figureheight]{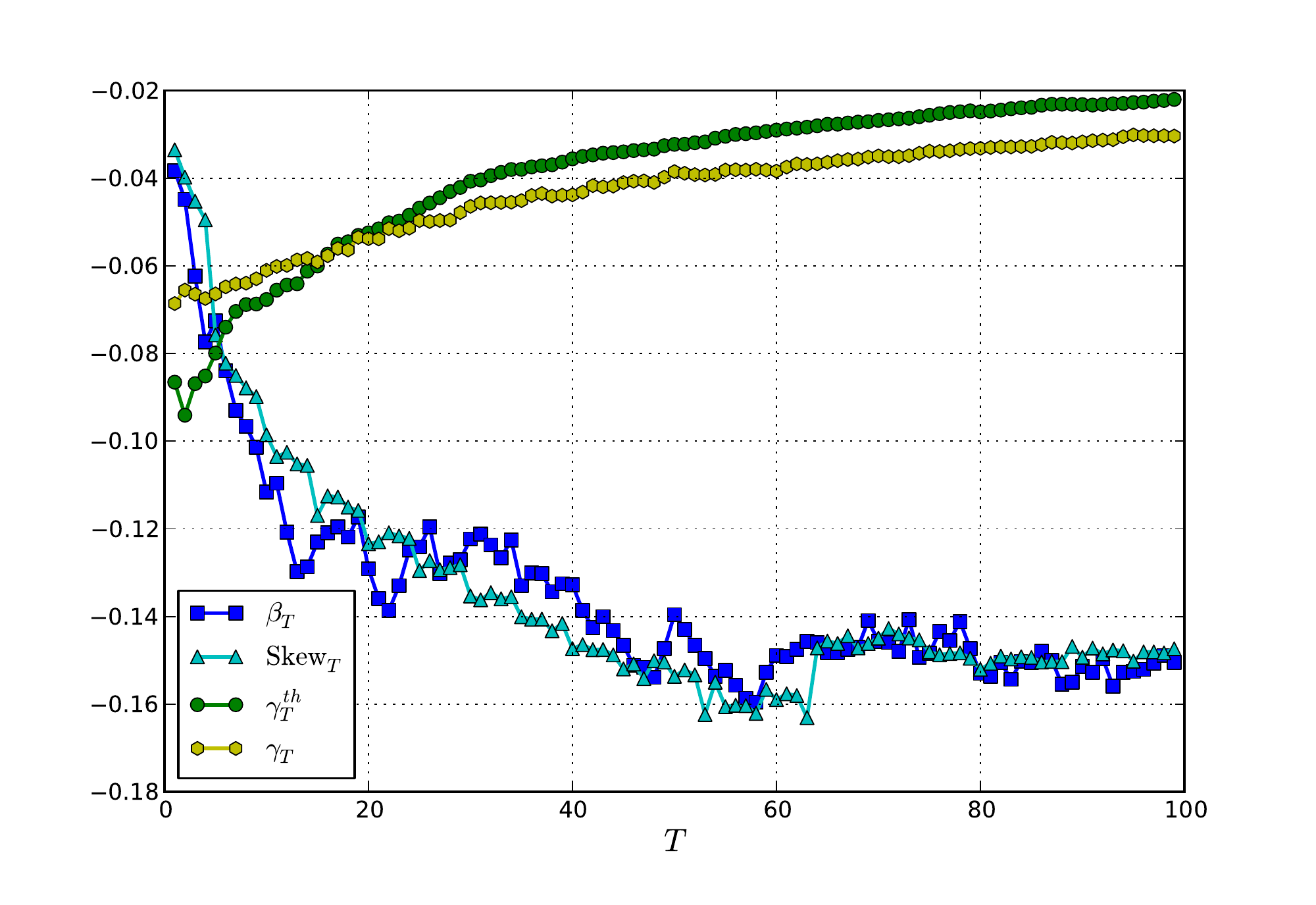}\\
	\caption{Comparison between theoretical prediction of the unconditional values of $\mathrm{Skew}_T$ and of $\gamma_T$, and their estimation based on option data. The left panel shows the result for S\&P 500 whereas the right panel is for DAX.}
\end{center}	
\end{figure}

\begin{figure}[tbph]
\begin{center}
	\includegraphics[width = 0.6\figurewidth,height = 0.6\figureheight]{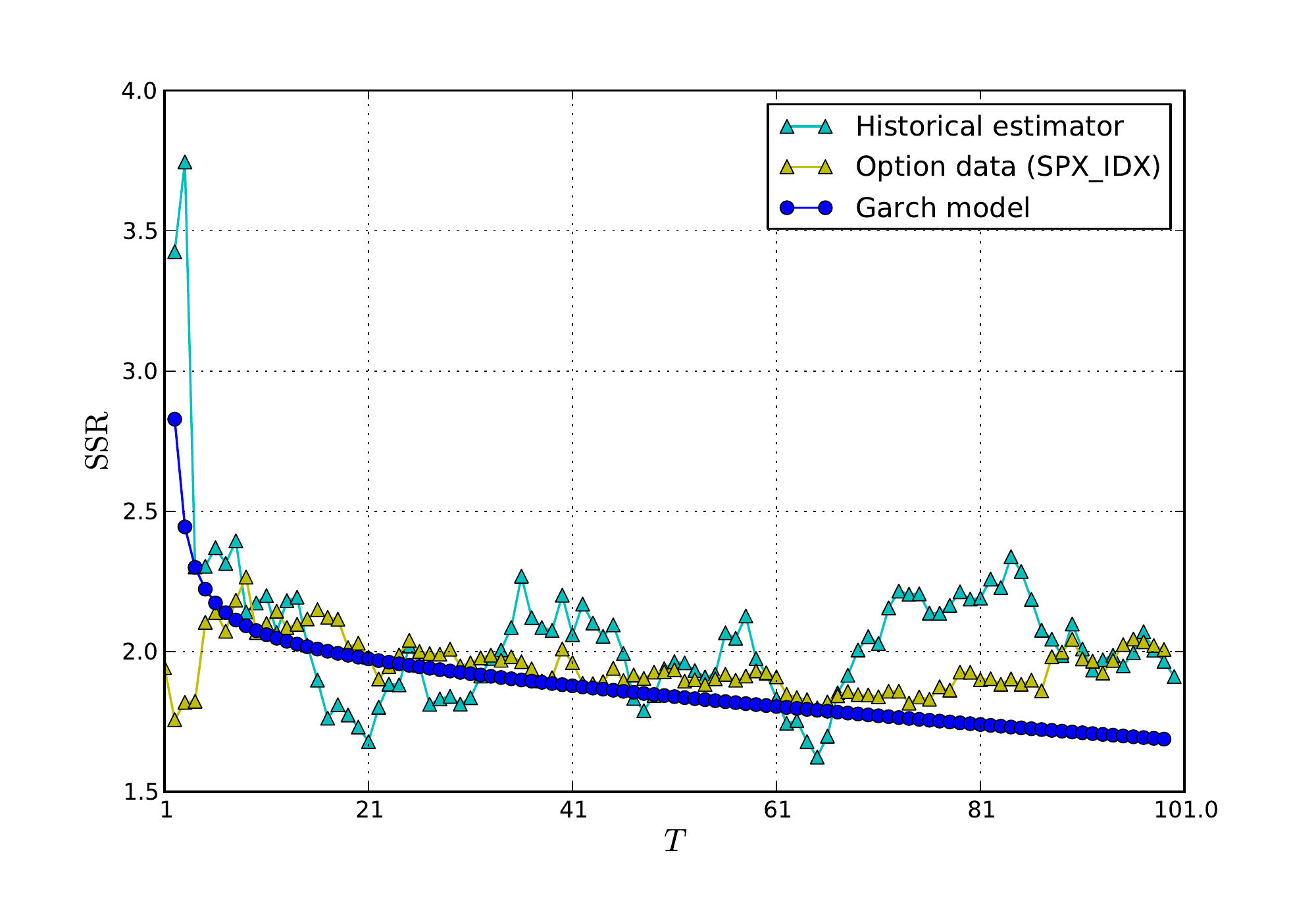}
	\includegraphics[width = 0.6\figurewidth,height = 0.6\figureheight]{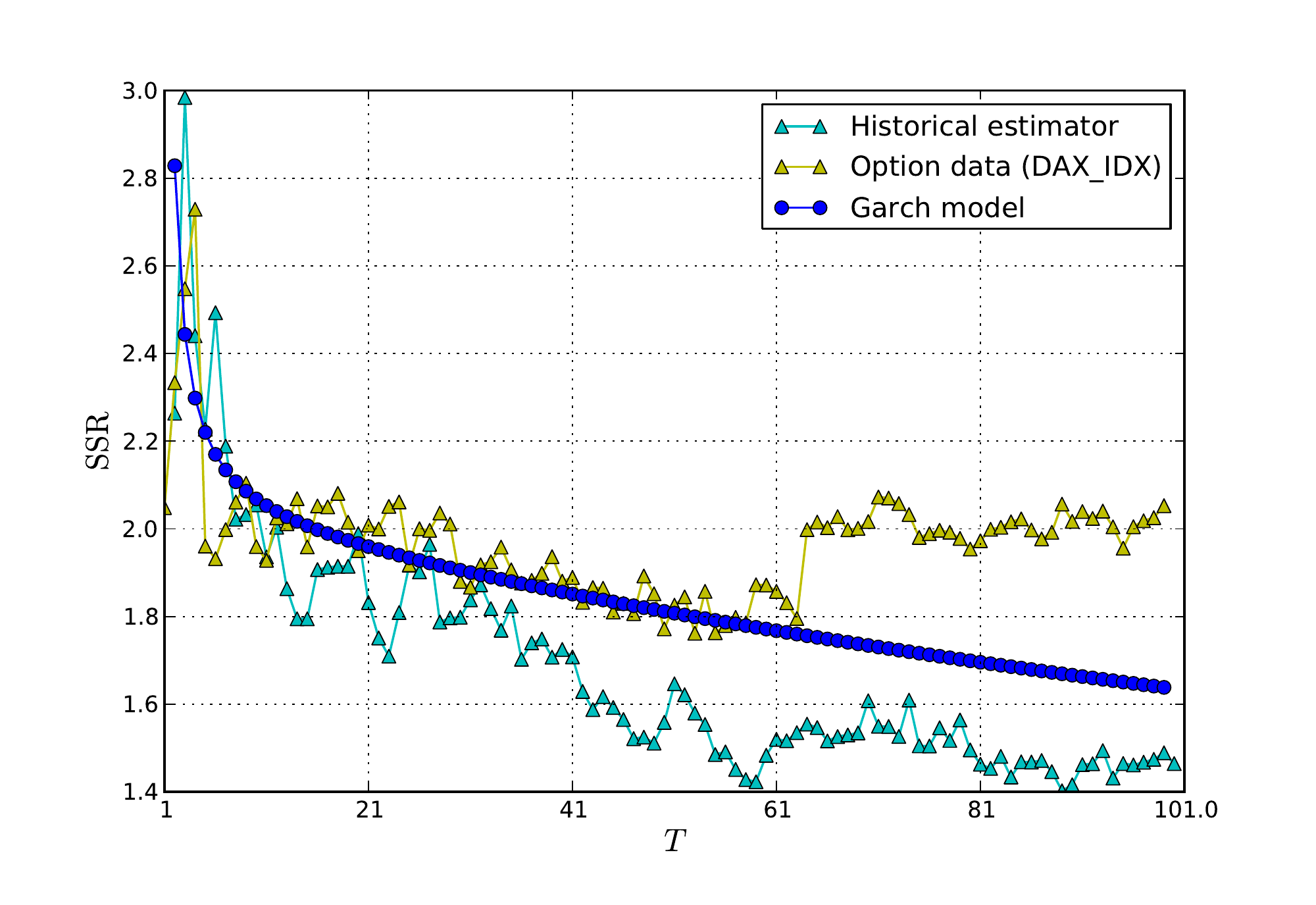}\\	
	\caption{Comparison between the real SSR estimated with option data as $\gamma_T \sqrt{T} /\mathrm{Skew}_T$ and the theoretical prediction based on 
	historical data, $\gamma_T^{th.}\sqrt{T}/\beta_T$. The left panel shows the result for S\&P 500 whereas the right panel is for DAX.}
\end{center}	
\end{figure}

\begin{figure}[tbph]
\begin{center}
	\includegraphics[width = 0.6\figurewidth,height = 0.6\figureheight]{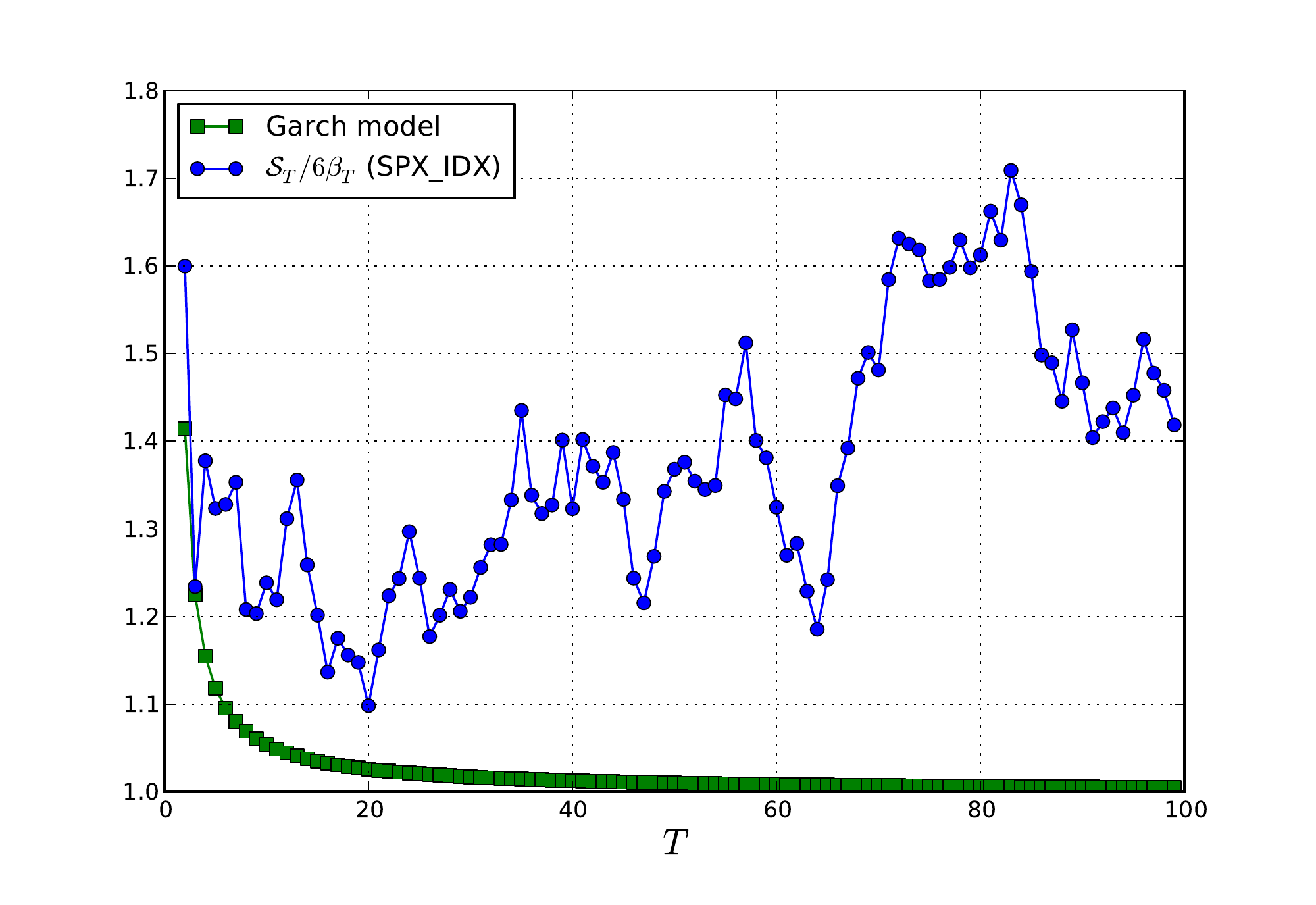}
	\includegraphics[width = 0.6\figurewidth,height = 0.6\figureheight]{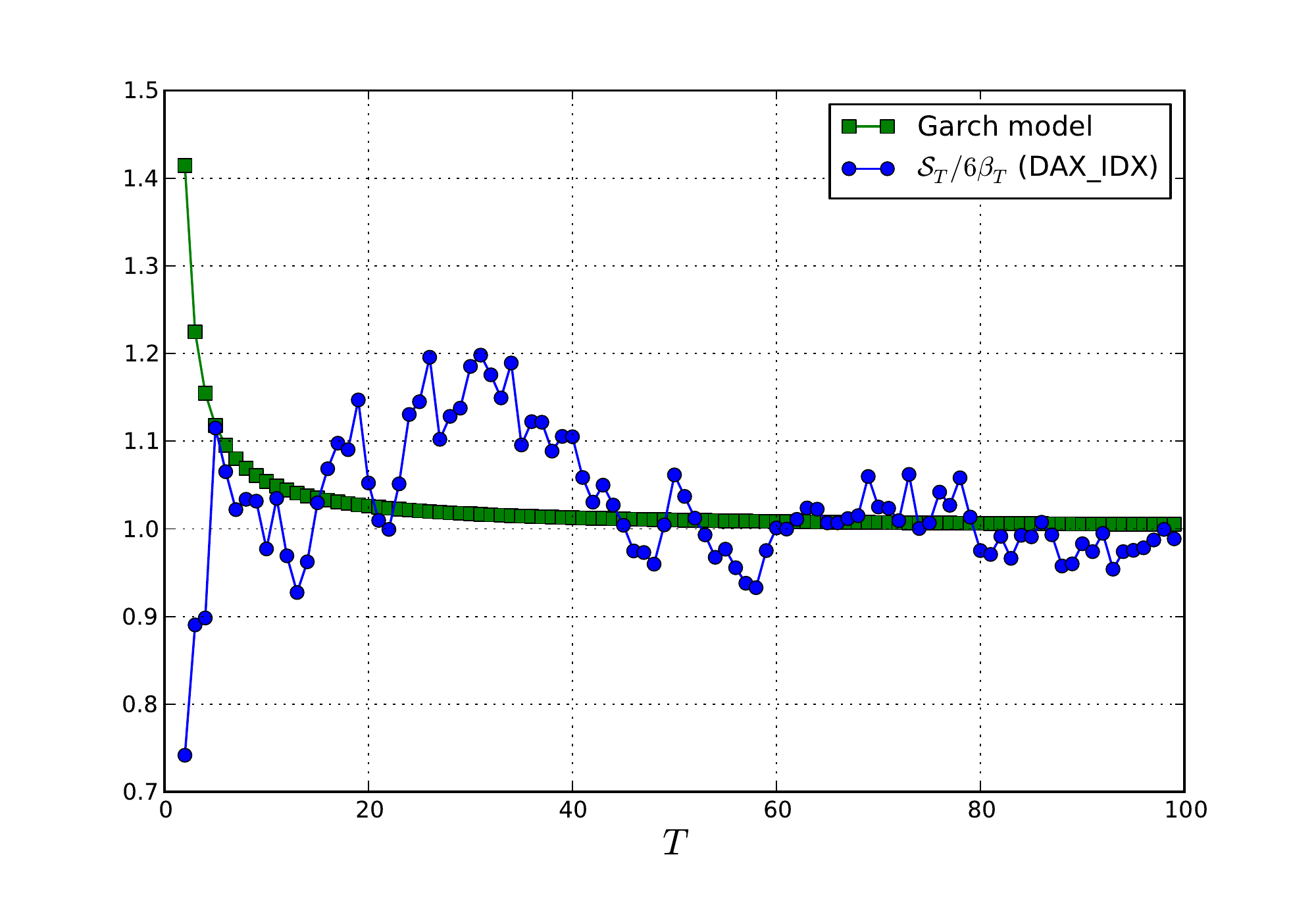}\\
	\caption{Comparison between $(\mathcal{S}_T/6)/\mathrm{Skew}_T$ estimated with S\&P 500 data and the prediction of GARCH model for S\&P 500 (left panel) and DAX (right panel).}
\end{center}	
\end{figure}

\section{Conclusion}

In this paper, we have revisited the problem of the dynamics of smiles, as envisaged in \cite{Ber2,CBP,CBP-E}, which consists in relating 
the ``implied leverage'' (i.e. the correlation of the at-the-money volatility with the returns of the underlying) and the skew of the option
smile. As noticed by Bergomi \cite{Ber2}, the ratio between these two quantities, dubbed the ``Skew-Stickiness Ratio'' (SSR), saturates to the value $2$ for 
linear models in the limit of small maturities, and converges to $1$ for long maturities, the latter value corresponding to the well-known ``Sticky 
Strike'' rule-of-thumb used by market makers. We have shown that for more general, non-linear models (such as the asymmetric GARCH model), Bergomi's 
result must be modified, and can be larger than $2$ for small maturities. The discrepancy comes from the fact that the volatility skew is, 
in general, {\it different} from the skewness of the underlying, 
as is found using either a cumulant expansion or a vol-of-vol expansion for linear models. The correct skew is rather given by a low-moment estimate of 
asymmetry, namely the difference between the probability of negative returns and the probability of positive returns (multiplied by $\sqrt{\pi/2}$). 
We compare our theory with empirical results, using data both from option markets and from the underlying price series, for the S\&P500 and the DAX.
We find, among other things, that although both the implied leverage $\gamma_T$ and the skew $\text{Skew}_T$ appear to be too strong on option markets (in particulr
for the S\&P500), their
ratio is well explained by the theory. We observe that the SSR clearly becomes larger than $2$ for small maturities. The asymmetric GARCH model, calibrated
on historical data, explains well the values of the SSR, but fails to reproduce accurately the different measures of skewness, in particular for the S\&P 500. 
The inadequacy of the asymmetric GARCH model to account for all the properties of options smiles was also noted in \cite{DVCB}.

It would also be quite interesting to extend our study to establish analoguous relations between the curvature of the smiles and measures of kurtosis \cite{DVCB}, 
and test them on data as well. 

We thank S. Ciliberti and L. De Leo for interesting discussions on these topics.

\section{Appendix: Proofs}

\subsection{Smile formula}

We prove the smile formula. By differentiating (\ref{definition}), we get the following decomposition at order 1 in $\nu$:
\begin{equation}
v_u^u = v_1^u+\nu \xi_u^{u,1}  
\end{equation}
where $v_1^u$ is the forward curve at time $0$ and $\xi_i^{u,1}$ is given by the following expression ($u \geq 2$):
\begin{equation}
\xi_u^{u,1} =  \sum_{j=1}^{u-1}  \lambda_j^u f(\epsilon_j) 
\end{equation}
Therefore, we get the following at order 1 in $\nu$:
\begin{equation} 
\sigma_u = \sqrt{v_1^u} + \frac{\nu}{2 \sqrt{v_1^u}}  \xi_u^{u,1}  
\end{equation}
We set the following:
\begin{equation}
V_T := \sum_{i=1}^T  v_1^i,  \quad N= \sum_{i=1}^T   \sqrt{v_1^i} \epsilon_i  -\frac{1}{2}  \sum_{i=1}^T  v_1^i
\end{equation}
and:
\begin{equation}
\tilde{N}= \frac{1}{2} \sum_{i=1}^T  \frac{ \xi_u^{u,1} }{ \sqrt{v_1^u}} - \frac{1}{2} \sum_{i=1}^T  \xi_u^{u,1} 
\end{equation}

Therefore, we have $\ln S_T= N + \nu \tilde{N}$
We introduce the function:
\begin{equation}
F(\nu)= E\left[ (S e ^{N + \nu \tilde{N}  }-K)_+   \right]
\end{equation}
Note that we get the following expression for the derivative:
\begin{equation}
F'(0)= S E\left[ \tilde{N} e^{N} 1_{N > \ln \frac{K}{S}}   \right]
\end{equation}
We get the following:
\begin{align*}
S E\left[ \tilde{N} e^{N} 1_{N > \ln \frac{K}{S}}   \right] &= \frac{S}{2}  \sum_{i=1}^T  \frac{1}{\sqrt{v_1^i}}  E\left[  \xi_i^{i,1} \epsilon_i   
e^{N} 1_{N > \ln \frac{K}{S}}   \right] - \frac{S}{2}  \sum_{i=1}^T   E\left[  \xi_i^{i,1}  e^{N} 1_{N > \ln \frac{K}{S}}   \right] \\
& = \frac{K}{2}  \sum_{i=1}^T E\left[ \xi_i^{i,1} \left | N= \ln \frac{K}{S}\right. \right] \frac{e^{-(\ln K/S+V_T/2)^2/2V_T}}{\sqrt{2 \pi V_T}}  \\
& = \frac{K}{2}  \sum_{i=1}^T \sum_{j=1}^{i-1} \lambda_j^i  E\left[ f(\epsilon_j) \left | N= \ln \frac{K}{S}\right.\right] \frac{e^{-(\ln K/S+V_T/2)^2/2V_T}}
{\sqrt{2 \pi V_T}}  \\
& = \frac{K}{2}  \sum_{i=1}^T \sum_{j=1}^{i-1} \lambda_j^i  E\left[ f(\epsilon_j) \left | N= \ln \frac{K}{S}\right.\right] 
\frac{e^{-(\ln K/S+V_T/2)^2/2V_T}}{\sqrt{2 \pi V_T}}  \\
\end{align*}
Now, recall that we have the decomposition $\epsilon_j= \frac{\sqrt{v_1^j}}{V_T} (N+\frac{V_T}{2}) +Y_j$ and thus we get:
\begin{equation}
S E\left[ \tilde{N} e^{N} 1_{N > \ln \frac{K}{S}}   \right] = \frac{K}{2}  \sum_{i=1}^T \sum_{j=1}^{i-1} \lambda_j^i  E\left[ f( \frac{\sqrt{v_1^j}}{V_T}(  \ln \frac{K}{S} +\frac{V_T}{2} ) +Y_j)\right] \frac{e^{-(\ln K/S+V_T/2)^2/2V_T}}{\sqrt{2 \pi V_T}}
\end{equation}

Now, by using the vega (\ref{vega}) below, we get:
\begin{equation}
 \delta V_T = \nu  \sum_{i=1}^T \sum_{j=1}^{i-1} \lambda_j^i  E\left[ f( \frac{\sqrt{v_1^j}}{V_T}\left(  \ln \frac{K}{S} +\frac{V_T}{2} \right) +Y_j)\right]
\end{equation}
Since $\delta \sigma_{ATM,T} = \frac{\delta V_T}{2 \sqrt{V_T T}}$, we get the smile formula.

\subsection{Skewness}
We get the following for the skewness $\mathcal{S}_T$ at order 1 in $\nu$:
\begin{align*}
\frac{\mathcal{S}_T} {6}  & =  \frac{1}{2} \frac{\sum_{i=2}^T \sum_{j=1}^{i-1}E[r_j \sigma_i^2]}{V_T^{3/2}}  \\
& =  \frac{\nu}{2}   \frac{\sum_{i=2}^T \sum_{j=1}^{i-1}  \sqrt{v_1^j} E[\epsilon_j \xi_i^{i,1}]}{V_T^{3/2}}   \\
& =  \frac{\nu}{2}   \frac{\sum_{i=2}^T \sum_{j=1}^{i-1}  \sqrt{v_1^j} \lambda_j^i }{V_T^{3/2}}  E[f'(\epsilon)]   \\
\end{align*}
\hspace{10 cm}

\subsection{Greeks}
 
\begin{lemma}[Greeks]

If we set $v \equiv V_T$ and:
\begin{equation}
\mathcal{{\rm BS}}(S,v)=S N\left(\frac{\ln(S/K)+\frac{1}{2}v}{\sqrt{v}}\right)-K N\left(\frac{\ln(S/K)-\frac{1}{2}v}{\sqrt{v}}\right)
\end{equation}
we get the following expression for the derivatives:
\begin{equation}
\frac{\partial \mathcal{{\rm BS}}(S,v)}{\partial S}=N\left(\frac{\ln(S/K)+\frac{1}{2}v}{\sqrt{v}}\right),
\end{equation}
\begin{equation}\label{vega}
\frac{\partial \mathcal{{\rm BS}}(S,v)}{\partial v}=\frac{S}{2\sqrt{v}}N'\left(\frac{\ln(S/K)+\frac{1}{2}v}{\sqrt{v}}\right)=\frac{S}{2\sqrt{2 \pi }\sqrt{v}}e^{-\frac{(\ln(S/K)+\frac{1}{2}v)^2}{2 v}}
\end{equation}
and
\begin{equation}
\frac{\partial^2 \mathcal{{\rm BS}}(S,v)}{\partial v^2}=-\frac{S}{4\sqrt{2 \pi } v^{3/2}}e^{-\frac{(\ln(S/K)+\frac{1}{2}v)^2}{2 v}}+\frac{S}{4\sqrt{2 \pi } v^{5/2}}( (\ln S/K)^2-v^2/4 )e^{-\frac{(\ln(S/K)+\frac{1}{2}v)^2}{2 v}}
\end{equation}
Finally,
\begin{equation}
\frac{\partial^{2} \mathcal{{\rm BS}}(S,v)}{\partial S \partial v}= \left(\frac{1}{4 \sqrt{v}}-\frac{\ln (S/K)}{2 v^{3/2}}\right)\frac{1}{\sqrt{2 \pi }}e^{-\frac{(\ln(S/K)+\frac{1}{2}v)^2}{2 v}}
\end{equation}

\end{lemma}

\end{document}